\documentclass[11pt,a4paper]{article}
\usepackage[hyperref]{acl2019}
\usepackage{times}
\usepackage{latexsym}

\usepackage[utf8]{inputenc} 
\usepackage[T1]{fontenc}    
\usepackage{hyperref}       
\usepackage{url}            
\usepackage{booktabs}       
\usepackage{amsfonts}       
\usepackage{graphicx}       
\usepackage{url}

\aclfinalcopy 

\title{How near-duplicate detection improves editors' \\ and authors' publishing experience}

\author{Yury Kashnitsky \\
  Elsevier, The Netherlands \\
  \texttt{y.kashnitskiy@elsevier.com}		 \\\And
  Vaishnavi Kandala\\
  Elsevier, The Netherlands \\
  \texttt{v.kandala@elsevier.com}			 \\\AND
  Egbert van Wezenbeek \\
  Elsevier, The Netherlands \\
  \texttt{e.wezenbeek@elsevier.com}		 \\\And
  IJsbrand Jan Aalbersberg \\
  Elsevier, The Netherlands \\
  \texttt{ij.j.aalbersberg@elsevier.com }	\\\AND
  Catriona Fennell \\
  Elsevier, The Netherlands \\
  \texttt{c.fennell@elsevier.com} 			\\\And
  Georgios Tsatsaronis \\
  Elsevier, The Netherlands \\
  \texttt{g.tsatsaronis@elsevier.com} }

\begin{document}
\maketitle

\begin{abstract}

We describe a system that helps identify manuscripts submitted to multiple journals at the same time. Also, we discuss potential applications of the near-duplicate detection technology when run with manuscript text content, including identification of simultaneous submissions, prevention of duplicate published articles, and improving article transfer service.


\end{abstract}


\section{Introduction}
\label{sec:intro}
Slow editorial processes and lack of communication around manuscript statuses can cause authors to submit to multiple journals in parallel. Publishers want to prevent processing such simultaneous submissions for various reasons including reputational concerns, legal ramifications, and to prevent overloading peer reviewers \cite{bmj_simult_subs,Wager2009}. A simple approach to the problem is to use manuscript content only (titles, authors, keywords, abstracts) to run near-duplicate detection, e.g. with Locality Sensitive Hashing \cite{leskovec}.


\section{Analysis of near-duplicate manuscripts}
\label{sec:analysis}

In this section, we describe the outcomes of a near-duplicate analysis that we did with manuscripts submitted to Elsevier journals. 

\textbf{Technical and methodological details:}

\begin{itemize}
\setlength\itemsep{-0.5em}
\item Following \cite{leskovec}, we say that two pieces of text are near-duplicates if the Jaccard similarity of their shingle sets exceeds some threshold, 0.8 in our case
\item We collected 4.6 mln. manuscripts submitted to >2K Elsevier journals between January 2018 and October 2020. 
\item We used titles and abstracts to run MinHash LSH for near-duplicate detection
\item For analysis, we implemented MinHash LSH with PySpark, the MLlib MinHash LSH implementation \footnote{\url{http://spark.apache.org/docs/2.2.0/api/python/pyspark.ml.html}} can serve as a reference 
\item The technique is limited to the analysis of lexical text similarity only, i.e. semantic similarity (paraphrasing, synonym replacement, etc.) can not be captured with LSH.
\end{itemize} 

\textbf{Insights and outcomes:}

\begin{itemize}
\setlength\itemsep{-0.5em}
\item \textbf{25\%} of manuscripts were found to have a near-duplicate submitted earlier
\item Over 90\% of near-duplicate pairs of manuscript titles are those where one manuscript is a \textit{resubmission} of another, meaning that it is submitted following a rejection of a similar manuscript
\item  \textbf{2.5\%} of manuscripts were found to be under consideration simultaneously at more than one journal.
\end{itemize} 


\section{Applications of manuscript near-duplicate detection}
\label{sec:applications}

Having developed a technology for large-scale detection of near-duplicate manuscripts, we see several applications of this technology for various use cases across Elsevier: catching simultaneous submissions, prevention of duplicate published article, and improving article transfer service.

\subsection{Catching simultaneous submissions}
\label{sec:simult_subs}

With over 2 mln. manuscripts submitted to Elsevier journals annually, and ~2.5\% of them being simultaneous submissions, flagging such cases to editors makes sense for the sake of saved reviewers' time. Having productionized the solution, we estimate we would be able to prevent reviewers from working on 50K redundant article reviews.

Preventing simultaneous submissions also ensures that authors get fair credit for their work which contributes to trust in the academic record \cite{Errami2008}. It can also help mitigate the problem of biased meta-analyses where duplicate studies are inadvertently included more than once \cite{Fairfield2017}.

\subsection{Prevention of duplicate published articles}
\label{sec:published_retractions}
Article retraction is a very undesired publication outcome for any publisher. One of the reasons for retractions is a simultaneous submission not caught on time. Although such cases are rare (we analyzed ~300 of them), they often lead to reputational ramifications not only for authors but for publishers as well. With near-duplicate detection technique in place, we can prevent some of these retractions.

\subsection{Improving article transfer service}
\label{sec:ats}

\begin{figure}[!t]
  \includegraphics[width=\linewidth]{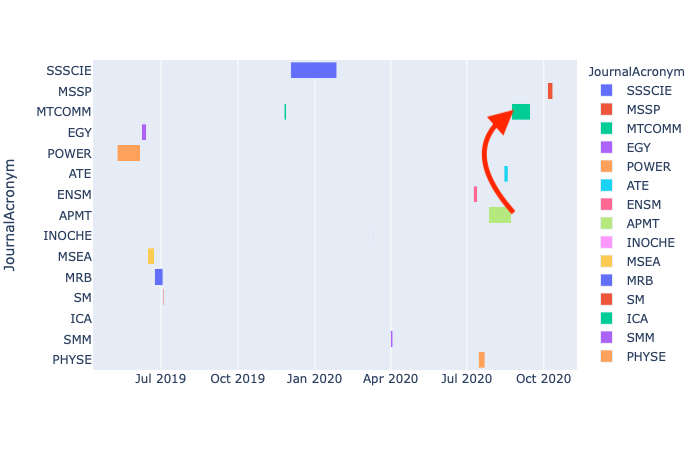}
  \caption{A manuscript journey and a bad article transfer. Axis $X$ stands for time, axis $Y$ stands for journals.}
  \label{fig_manu_journey}
\end{figure}

One possible optimization of the editorial process would be the improvement of the article transfer service which can be done if we avoid transfers to a journal that had already processed the given manuscript (or its near-duplicates) in the past. With near-duplicate detection technique, we can build such ``manuscript journeys'' that track journals that had processed a given manuscript (or its near-duplicates), see Fig. \ref{fig_manu_journey}. A red arrow corresponds to such a ``bad'' transfer. With a preliminary analysis, we found that ~1\% of transfers can be improved this way.
 
Also, we can collect statistics from past resubmissions and use historic data on final successful publication to make better (higher chance of acceptance) recommendations for transfers and thus improve author experience.

 
\section{Conclusions}
\label{sec:future}

We see that a simple near-duplicate text detection technique, if implemented efficiently and scaled, has attached to it many important use cases leading to better and safer editorial processes, an increased author experience and a larger degree of integrity in the publication process. The technique can further be improved to account for semantic difference of the textual content like paraphrasing or changing word order.

\bibliography{21_02_CRI_research_integrity_conf_resub}
\bibliographystyle{acl_natbib}

\end{document}